\documentclass[fleqn,twoside]{article}
\usepackage{espcrc2,epsf,epsfig,psfrag,graphicx} 

\newcommand{\be}{\begin{equation}}
\newcommand{\ee}{\end{equation}}
\renewcommand{\thefootnote}{\fnsymbol{footnote}}

\title{Solar Neutrinos: Interpretation of Results}
\vspace{0.5cm}

\author{A. Yu. Smirnov\address{
The Abdus Salam International Centre for Theoretical Physics,  
I-34100 Trieste, Italy,\\ 
Institute for Nuclear Research of Russian Academy 
of Sciences, Moscow 117312, Russia}\thanks{
Invited talk given at the XXth International Conference on
Neutrino Physics and Astrophysics, ``Neutrino 2002", 
Munich, Germany, May 25-30, 2002. }}

\begin{document}

\begin{abstract}

Recent SNO results give strong evidence  that the 
solar neutrinos undergo the flavor conversion. 
Main issue now is  the identification of  the mechanism
of conversion. The LMA MSW solution with $\Delta m^2 = (5 - 7) \cdot
10^{-5}$ eV$^2$, $\tan^2 \theta = 0.35 - 0.45$
looks rather plausible: it fits well the experimental  data and our new
theoretical prejudices. In the LMA case, KamLAND should see  
(0.5 - 0.7) reduced signal. 
VAC-QVO and  LOW  are accepted  at about
$3\sigma$-level.  The SMA solution is practically excluded. 
No sub-leading effects produced by  
$U_{e3}$ and admixture of sterile neutrino have been found. 
The fit becomes worse with increase of $U_{e3}$ (for LMA) and the
$\nu_s$ admixture. Still (30 - 50)\% presence  of the sterile neutrino
is allowed.  Solutions based on the neutrino spin-flip in the magnetic
fields of the Sun  as  well as on non-standard neutrino interactions give
good fit of the data. If KamLAND confirms LMA MSW, the 
spin-flip and non-standard interactions can be considered (and will be 
searched for) as sub-leading effects. 

\end{abstract}
\maketitle

\renewcommand{\thefootnote}{\arabic{footnote}}
\setcounter{footnote}{0}
\renewcommand{\baselinestretch}{0.9}

\section{INTRODUCTION}

The SNO results \cite{sno,sno-nc,sno-dn,sno-how}  have led to 
breakthrough in the solar neutrino studies. Main conclusion is 
that solar neutrinos undergo the flavor conversion
\be  
\nu_e \rightarrow \nu_{\mu}, \nu_{\tau}~~{\rm or/and} ~~ \bar\nu_{\mu},
\bar\nu_{\tau}.  
\ee
The key word is ``appearance": the appearance of 
the muon and tau neutrinos in the solar neutrino flux. 
Moreover,  it seems, $\nu_{\mu}$ and  $\nu_{\tau}$ compose larger part of
the solar neutrino flux at high energies. 
These results further  confirm earlier indications 
of $\nu_{\mu}, \nu_{\tau}$ appearance 
from comparison of signals in the Homestake~\cite{Cl}, 
and Kamiokande / SuperKamiokande (SK)~\cite{SK} experiments, and later  
from comparison of fluxes determined by CC SNO results~\cite{sno}  and 
the $\nu e-$scattering events at Super-Kamiokande~\cite{SK,sk-nu}. 
 
The main issue now is to identify the {\it mechanism} of neutrino
conversion.

This review is  written about two months after the conference.    
During this time 
a number of new studies of the solar neutrino data have been published. 
Some new points have been realized but
general picture and conclusions have not been changed. 
I will present an updated analysis, including the latest SAGE~\cite{sage}
and GNO~\cite{gno} results.  

This  review is written  several weeks (months?) before 
announcement of the  KamLAND result which may put final ``dot" in the
long story  of solar neutrino problem and make substantial
part of the discussion below to be irrelevant.  

\section{PROFILE OF THE EFFECT} 

Identifying  the mechanism of
conversion one  looks for signatures 
in the energy and time dependence of observables.

\subsection{The energy profile} 

Present data allow to reconstruct the energy dependence of the effect 
without  reference to  certain mechanism of conversion
\cite{BL,FLMPa,barger-a}. 
The only assumption is that sterile neutrinos, if exist, do not
participate
in the solar neutrino conversion. 
Schematically the procedure of reconstruction can be performed
in the following three steps (see, {\it e.g.},~\cite{BB,Ber} for
earlier analysis). 

\noindent
1).``SNO NC/CC and SK/SNO". Using the SK and SNO data one can get the
averaged survival probability
above $E > 5$  MeV independently on the SSM predictions of fluxes. 
According to  SNO and SK there is no substantial dependence of the
survival probability on energy. 
Therefore  the  average probability $P_{ee}$ can be  immediately 
obtained  from the ratio of the CC and  NC event rates:    
\be
P_{ee} = \frac{CC}{NC} = 0.345^{+ 0.045}_{- 0.040}.  
\label{nc/cc}
\ee              

\noindent
2).``Cl - SK/SNO over SSM". The SK and SNO data  allow to evaluate
original boron  neutrino flux, and consequently, to calculate the
contribution of this flux to
the $Ar$-production rate $Q_{Ar}$ . Subtracting this contribution from the
rate measured in Homestake  one finds the contribution
of fluxes of intermediate energies (0.8 - 1.5) MeV to $Q_{Ar}$. Then using
the SSM predictions for the fluxes of $Be$, $pep$ and $CNO$-cycle
neutrinos one gets the survival probability
$P_{ee}(0.8 - 1.5 {\rm MeV})$.

\noindent   
3).``Ga - Cl - SK/SNO over SSM". Using  $P_{ee}(0.8 - 1.5
{\rm MeV})$ found in the previous step 
and  SSM fluxes of $Be$ , $pep$ and $CNO$-cycle neutrinos
one can estimate contributions of these fluxes to the $Ge$-production
rate,  $Q_{Ge}$.
Subtracting this contributions as well as the contribution of
boron neutrinos from $Q_{Ge}$ measured by 
SAGE and GNO/GALLEX  one gets the contribution 
of $pp$-neutrinos to $Q_{Ge}$. Then using the SSM
prediction for the $pp$-neutrino flux one finds 
$P_{ee}(0.2 - 0.4 {\rm MeV})$.


\begin{figure}[ht]
\centering\leavevmode
\epsfxsize=.4\hsize
\includegraphics[width=0.40\textwidth]{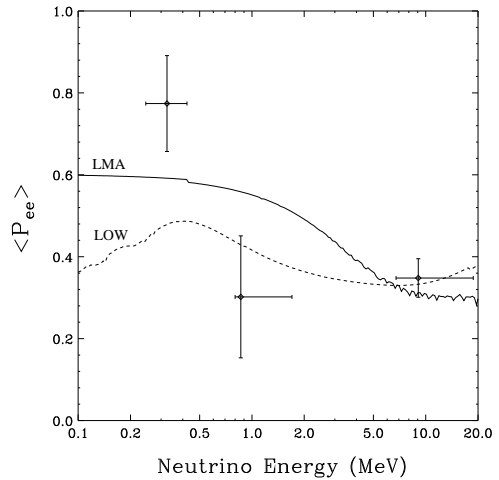}
\vskip -1cm
\caption{
The profile of the effect. Shown are the reconstructed values of the 
survival probability in different energy ranges. The lines correspond to
the survival probability for  the LMA and LOW solutions; (from
\cite{barger-a}). 
}
\label{fi1}
\end{figure}     

The result of such a type of  analysis is shown in fig.~\ref{fi1} 
taken from~\cite{barger-a}.  The best profile 
(which corresponds to the central values of reconstructed $P_{ee}$) 
would have constant
$P_{ee}$ at high energies, stronger suppression at the intermediate
energies and weak suppression at low energies.
The profile can not be reproduced by LMA or LOW solutions.
Although these solutions are in agreement with reconstructed profile
within the error bars. Notice that
the errors of reconstructed values of $P_{ee}$ at different  energies
strongly correlate: higher  value of $P_{ee}$
at the intermediate energies corresponds to
lower value of $P_{ee}$ at low energies. 
The reconstructed profile can be reproduced precisely by the resonance
spin-flavor precession.
 
\subsection{Time variations} 

\noindent
1). 11 years variations related to the solar activity cycle:
no variations have been found in SK, SAGE and GALLEX/GNO experiments.
The cause  of possible changes of the  Homestake rate still unclear.

\noindent
2). Seasonal variations: SK data are in agreement with geometrical 
(eccentricity) effect. No statistically significant variations
of $Ge$-production rate have been found~\cite{sage,gno}.

\noindent
3). Monthly variations which  can be related to rotation of the Sun: 
it is argued that gallium and  Clorine results
show time variations with the several weeks period \cite{Stur}.
The claim is based on  direct analysis of time dependence of 
experimental signals as well as on bi-normal distribution of number of
bins with a given rate in GALLEX/GNO and SAGE. It should be
stressed, however,  that GNO results alone do not show bi-normal
distribution~\cite{gno}.
 
\noindent
4). Diurnal variations: the day-night asymmetries measured by
SK and SNO(CC) experiments are
\be
- A^{ES}_{DN} =  2.1 \pm 2.0 ~^{+1.3}_{-1.2}  \%  , 
\label{asymSK}
\ee
\be
- A^{CC}_{DN} = 7.0 \pm 4.9 ~^{+1.5}_{-1.4}  \%  .        
\label{asymSNO}
\ee
The later is obtained under constraint that total flux has no D-N
asymmetry~\cite{sno-dn}.   
Although the results are not statistically significant, they have
the expected sign and values.
The difference of asymmetries at SNO and SK is explained by
the dumping factor  for ES signal due to effect of $\nu_{\mu},
\nu_{\tau}$ \cite{BKS10}. 
For LMA and LOW solutions we get
\be
A_{DN}^{CC} = \left[1 + \frac{r}{(1 - r)P_{ee}}\right] A_{DN}^{ES},   
\ee
where $r \equiv \sigma(\nu_{\mu} e)/\sigma(\nu_{e} e)$. The effect is
significant (in spite of small $r$)
since the  $\nu_{\mu}/\nu_{\tau}$ flux is 2 - 4 times larger
than the $\nu_e-$ flux.
Taking $P_{ee}$  from (\ref{nc/cc}) we get for LMA: 
$A_{DN}^{CC} = 1.53 A_{DN}^{ES}$, in agreement (within 1$\sigma$) 
with the experimental results.

\section{GLOBAL FIT}
Implications of new SNO results have been studied
in~\cite{barger-a,bahcall-a,dproy,strumia,milano,pedro,cattadori,bari,valen-so}. 
Here I will present  our updated  analysis \cite{pedro} 
which includes, in particular, new GNO results.  

\subsection{Input} 

\noindent
{\it Data.} In the analysis we use 
(1) three rates: $Q_{Ar}$,
(Homestake experiment \cite{Cl}), $Q_{Ge}$, from 
SAGE \cite{sage}, and  combined $Q_{Ge}$ from GALLEX
and GNO \cite{gno};   (2) the zenith-spectra 
measured by Super-Kamiokande \cite{SK} during 1496 days~\cite{SK,sk-nu}; 
(3) the day and the night energy spectra 
of all events measured at SNO \cite{sno-how}. 
Altogether there are 81  data points. 

\noindent
{\it Neutrino Fluxes.} 
The boron neutrino flux  (the dimensionless quantity 
$f_B \equiv F_B/F_B^{SSM}$),  is treated as  free parameter. 
The SSM  boron neutrino flux is  
$F^{SSM}_B = 5.05 \cdot 10^{6}$ cm$^{-2}$ s$^{-1}$.  
All  other neutrino fluxes and their uncertainties 
are taken according to SSM BP2000~\cite{ssm}. 
For the $hep-$neutrino flux we use fixed value  
$F_{hep} = 9.3 \times 10^{3}$ cm$^{-2}$ s$^{-1}$ \cite{ssm,hepfl} .

In the case of two neutrinos  there are three 
fit parameters: $\Delta m^2$, $\tan^2\theta$, $f_B$, 
and therefore 81(data points) - 3 = 78 d.o.f. .

\subsection{ LMA MSW}
\null
\vskip 0.7cm
\begin{figure}[htb] 
\centering\leavevmode
\epsfxsize=0.6\hsize
\includegraphics[width=0.46\textwidth]{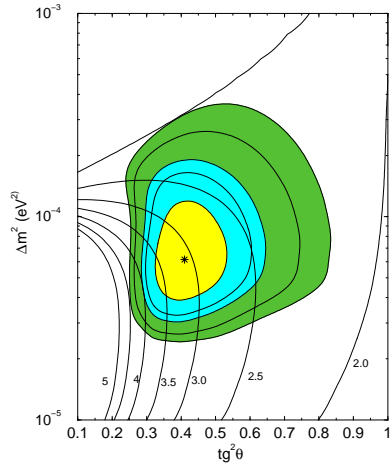}
\vskip -1.5cm
\caption{
The global  LMA MSW solution. The best fit point is marked by a star.  
The allowed regions   are shown at 1$\sigma$, 90\% C.L., 2$\sigma$, 
99\% C.L. and 3$\sigma$.  
Also shown are the lines of constant NC/CC ratio. 
}
\label{nccc_lma} 
\end{figure}

\begin{figure}[htb]
\epsfxsize=.6\hsize
\includegraphics[width=0.46\textwidth]{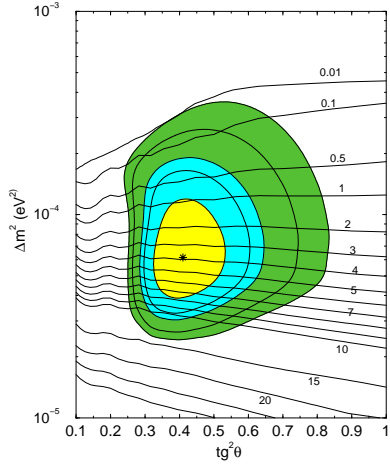}
\vskip -1.5cm
\caption{
Lines of constant day-night asymmetry of CC events.  
In the best fit point: $A_{DN}^{CC}=3.9\%$.
}
\label{adn_lma} 
\end{figure}

In the best fit point (see fig.~\ref{nccc_lma}) 
\be
\begin{array}{l}
\Delta m^2 =  6.15 \cdot 10^{-5}~ {\rm eV}^2, ~~~\tan^2 \theta = 0.41,\\ 
f_B = 1.05,
\end{array}
\label{bfparam}
\ee 
we get $\chi^2/dof = 65.2/78$.  In fig.~\ref{nccc_lma} we show
also the lines of constant ratio NC/CC. 
The best fit point corresponds to ${\rm NC/CC} = 3.15$. 
The best fit  values and $3\sigma$ intervals  
for $Ar-$ and  $Ge-$ production rates  equal 
\be
Q_{Ar} = 2.95~ ^{+0.40}_{-0.25}~{\rm SNU},~~ 
Q_{Ge} = 70.5~ ^{+13.5}_{-7.5}~{\rm SNU}.  
\ee 
The grid of predicted values of DN-asymmetry in CC is shown in
fig.~\ref{adn_lma}.

{\it How large is the large mixing?} The  upper limit on mixing angle is
controlled by the following observables: 
\be
\begin{array}{l} 
\frac{CC}{NC} \sim \sin^2\theta, ~~~ Q_{Ar} \sim f_B \sin^2\theta,\\ 
Q_{Ge} \sim 1 - \frac{1}{2}\sin^2 2\theta. 
\label{theta-up}
\end{array}  
\ee
The global fit gives  
\be
\tan^2 \theta <  0.84, ~~  99.73 \%  ~~ {\rm  C.L.} .\\
\label{theta-up}
\ee 
\null
\vskip 0.7cm  
There is a  significant spread of the bounds  obtained by different
groups: 0.55 \cite{sno-dn}, 0.64 \cite{barger-a}, 0.89 \cite{bahcall-a}  
$( 99.73 \%  ~~ {\rm  C.L.})$. 
In any case, we have  a strong evidence that solar neutrino mixing
significantly 
deviates from maximal value. The 
parameter~\cite{max} 
\be
\epsilon \equiv 1 - 2\sin^2\theta,  
\ee 
which characterizes the deviation of mixing from the maximal one, 
equals~\cite{pedro} 
\be
\epsilon  > 0.08,~~~(3 \sigma).
\ee 
That is, at $3 \sigma$: 
$\epsilon > \sin^2\theta_c$,  where $\theta_c$ is the Cabibbo angle. 
This result has important implications for the  
neutrinoless double beta decay searches, determination of the absolute mass
scale 
of neutrinos, theory of neutrino masses.    
Maximal mixing is accepted   
at $4 \sigma$ level in the range $\Delta m^2 = (7 - 10) \cdot 10^{-5}$
eV$^2$.  For maximal mixing we get: 
NC/CC $< 2~~ (-2.2 \sigma)$, $Q_{Ar} \sim 3.2~ {\rm SNU} ~~ (+ 2.8
\sigma)$, 
$Q_{Ge} \sim 64~{\rm SNU} ~~ (- 1.8 \sigma)$, $f_B = 0.85$. 

{\it How high is the high $\Delta m^2$?}  The upper bound on $\Delta m^2$ 
has important implications for measurements of $\Delta m^2$ itself, 
future LBL experiments, determination of 
the CP-violating phase, etc.. 
From the global fit we find 
\be
\Delta m^2 \leq 
3.6 \times 10^{-4} ~ {\rm eV}^2,  ~~99.73\% ~ {\rm C.L.} 
\label{up-on-dms}
\ee
which is  stronger than the 
CHOOZ bound~\cite{CHOOZ}.

\subsection{Physics of LMA}

Physical processes 
depend  on neutrino energy and values of oscillation
parameters (see {\it e.g.} \cite{lma99,zenith} for recent discussion).
According to the LMA MSW solution (with the parameters 
(\ref{bfparam})) neutrinos undergo the adiabatic
conversion inside the Sun. 

1). For high energy neutrinos ($E > 5$ MeV) the conversion  has a
character 
of non-oscillatory adiabatic transformation $\nu_e \rightarrow \nu_2$, 
with final survival probability $P_{ee} \approx \sin^2\theta$.
In  matter of the Earth $\nu_2$ oscillates leading to $\nu_e$ 
regeneration effect which  increases with energy  and can reach few per
cents (see, {\it e.g.}, \cite{zenith}). 

2). In  the intermediate energy range ($0.8 - 2$ MeV) 
there is an interplay of the
conversion and oscillations. The later are averaged out.

3). At low energies ($E < 0.5$ MeV) the effect is reduced to vacuum
oscillations with small matter corrections: 
$P^{(2)}_{ee} \rightarrow (1 -  0.5 \sin^2 2\theta_{12})$. 

\subsection{Pro and Contra}
1). LMA does not reproduce perfectly the energy profile of the effect.
LMA predicts larger $Q_{Ar}$ than the Homestake rate.
It is not excluded, however,  that Homestake has some unknown 
systematic error.

2). No one signature of the LMA has been observed with high statistical
significance: There is no  turn up of the spectrum at low energies, 
although,  the present sensitivity is not enough to observe the
effect. There is no significant regeneration effect.
At the same time both SK and SNO data give  some indications
of the day-night asymmetries which have correct sign and 
correct relative values.

3). The oscillation solution of the  atmospheric neutrino problem 
makes rather plausible interpretation of the solar neutrino results in
terms
of vacuum mixing and masses.
Also large/maximal mixing observed in atmospheric neutrinos give 
the precedent for large mixing in conversion of solar neutrinos.

4). In the case of LMA MSW the  hierarchy
of solar and atmospheric $\Delta m^2$ is the weakest  
which is natural in presence of large mixing.

\subsection{Predictions for KamLAND} 
\null
\vskip 0.7cm  
\begin{figure}[htb]
\epsfxsize=.6\hsize
\includegraphics[width=0.46\textwidth]{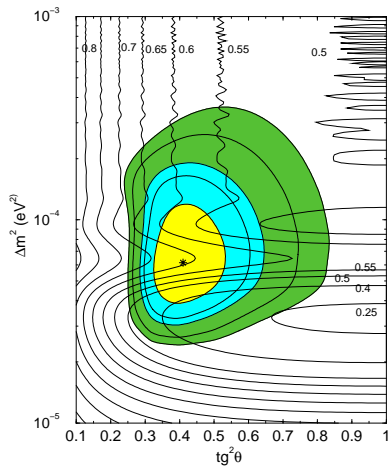}
\vskip -1.5cm
\caption{Lines of constant suppression of the rate at KamLAND, 
$R_{Kam}$, (figures at the curves).}
\label{kamt}
\end{figure}  

The discussion of {\it pro and contra}  LMA will be irrelevant after 
the KamLAND results~\cite{busenitz}. 
The effect of oscillation disappearance  can be characterized by  the
ratio, $R_{Kam}$, of the total number of events 
with visible energy  $T > 2.6$ MeV with and without oscillations. 
In fig.~\ref{kamt}  the contours of constant suppression factor $R_{Kam}$ 
are shown in the $\Delta m^2 - \tan^2\theta$ plot. 
Details of calculations can be found in~\cite{pedro}. 
In the best fit point we find   
$R_{Kam} = 0.65$,  the $1\sigma$ region is $R_{Kam} = 0.4 - 0.7$, 
and in $3\sigma$ region: $R_{Kam} = 0.25 - 0.73$. 
The value $R_{Kam} >  0.75$ will be problematic for the LMA.  
The distortion of the visible energy spectrum depends on 
$\Delta m^2$ strongly.  
The most profound effect of oscillations is  suppression of the rate at
high energies.  For instance, for $E \approx 5$ MeV 
and  b.f. value of $\Delta m^2$ the suppression 
is stronger  than 1/2.

\section{ALTERNATIVES}

Viable alternatives to LMA MSW 
can be divided in to  three categories, depending on mechanism of
conversion:
(1) LOW MSW and VO-QVO related to  transformations induced by 
neutrino masses and vacuum mixing. 
(2) Solutions based on neutrino spin-flip in the
magnetic fields of the Sun. Here there are two possibilities: 
the flip in the convective zone, and in the radiative
zone of the Sun. Furthermore, the effect of flip can have resonance
or non-resonance character. (3) Conversion induced by 
non-standard neutrino interactions. 
For all these solutions (but VO), the matter effects play crucial role.

\subsection{Mass and mixing solutions}
\noindent
{\it 1).  VAC-QVO next best?}.  In the best fit point  
\be
\begin{array}{l}
\Delta m^2 =  4.5 \cdot 10^{-10}~ {\rm eV}^2,
~~~\tan^2 \theta = 2.1,\\
f_B = 0.75, 
\end{array}
\label{bfparamvo}
\ee
we get $\chi^2(VAC) -  \chi^2(LMA) = 9.7$.    
The solution is accepted at $3\sigma$ level. 
It  appears in the dark side of the parameter space 
so that some matter effect is present. 
Features of this  solution ``discovered'' in 1998 
are:  absence of  any day-night asymmetry,   
low ($- 1.6\sigma$)  Boron neutrino flux,  
high    ($+ 2.7 \sigma$) Ar-production rate: 
$Q_{Ar} = 3.2$ SNU, low ($- 2.6\sigma$) ratio: NC/CC$= 1.86$.\\ 


\noindent
{\it 2) LOW starts to  disappear?}  In  the best fit point   
\be
\begin{array}{l}
\Delta m^2 =  0.93 \cdot 10^{-7}~ {\rm eV}^2,
~~~\tan^2 \theta = 0.64,\\ 
f_B = 0.91, 
\end{array}
\label{bfparamvo}
\ee 
we get  $\chi^2(LOW) -  \chi^2(LMA) = 12.4$
which is slightly  beyond the $3\sigma$ range. 
In other analyses, LOW does  appear at $3\sigma$ level. 
Inclusion of 
the SK data which contain information about zenith angle distribution 
(zenith spectra) worsens the fit (see comments in  
\cite{cattadori}).

The LOW solution gives rather poor fit of total rates. 
In the best fit point we get large 
($+ 2.1 \sigma$) $Ar-$production rate:  $Q_{Ar} = 3.2$ SNU 
and  $1.2 \sigma$ lower
$Ge-$production rate: $Q_{Ge} = 66.5$ SNU. 
The ratio NC/CC$= 2.35$ is in agreement with observations.  
For the day-night asymmetry of the CC-events we predict 
$A_{DN}^{CC} = - 3.5 \%$  and for ES  events: $A_{DN}^{CC} = - 2.7 \%$.

\noindent
{\it 3).  Any chance for SMA?} 
In  the best fit point  
\be
\begin{array}{l}
\Delta m^2 =  5.0 \cdot 10^{-6}~ {\rm eV}^2,
~~~\tan^2 \theta =  5.0 \cdot 10^{-4}, \\
f_B = 0.58,
\end{array}
\label{bfparamvo}
\ee
we get  $\chi^2(SMA) - \chi^2(LMA) = 34.5$. 
So, SMA is accepted at  $5.5\sigma$ only. Moreover, the 
solution requires about $3\sigma$ lower boron neutrino 
flux than in the SSM.   
It  predicts positive Day-Night asymmetry: 
$A_{DN}^{CC} =  0.93\%$. 
Very bad fit is due to 
latest SNO measurements of the day and night spectra. 
We find $NC/CC \approx 1/P_{ee} = 1.37$
which is substantially smaller than the observed quantity (\ref{nc/cc}).
Contribution of the $NC$ events 
is suppressed which leads to distortion of the energy spectrum of all 
events in comparison with observations. 
At the same time, SMA provides rather
good description of the SK data: the rate and spectra. 
(We find that $\chi^2$ increases weakly with 
$\tan^2 \theta$ up to $\tan^2 \theta = 1.5 \cdot 10^{-3}$,  where 
$\chi^2 \sim 105$.) 

The solution is practically  excluded unless some  systematic 
error will be found in the SNO or/and SK results.

\subsection{Pull-off diagrams.}

\begin{figure}[ht]
\centering\leavevmode
\epsfxsize=1\hsize
\epsfbox{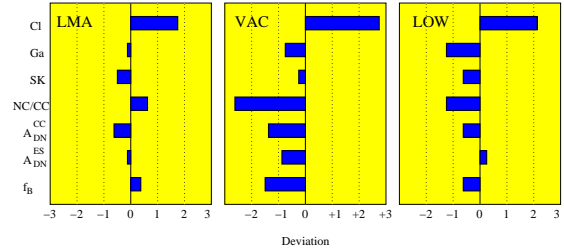}
\vspace{-1.5cm}
\caption{
Pull-off diagrams for the LMA, VAC and LOW solutions. 
The pull-offs  are expressed  in the units of 1 standard deviation,  
$1 \sigma$.}
\label{pulloff} 
\end{figure}


The pull-off diagrams (fig.~\ref{pulloff}, \cite{plamen,pedro}, see 
\cite{bari} for extensive discussion of the pull method)   
show deviations, $D_K$, of the predicted
value of observable $K$ in the best fit point  
from the central experimental value 
expressed in the $1\sigma$ unit, $\sigma_K$,:
\be
D_K \equiv \frac{K_{bf} - K_{exp}}{\sigma_K}, ~~~~ 
\label{pull}
\ee 
$K \equiv Q_{Ar},~Q_{Ge},$~ NC/CC,~ $R_{\nu e},~ A_{DN}^{SK},
A_{DN}^{CC}$.   
We take the experimental errors only: $\sigma_K = \sigma_K^{exp}$. 

According to  the pull-off diagram 
the LMA solution reproduces 
observables at $\sim 1\sigma$ or better. 
LOW and VAC  give  worse fit to the data.

\section{LEADING AND SUB-LEADING}

Identification of the
solution means, first of all, the identification of  the leading effect
in solar neutrinos. LMA MSW is rather plausible candidate.
Apart from the leading mechanism a number of  sub-leading effects
can be present. 

According to available data, 
$U_{e3}$, mixing of sterile neutrinos,  and neutrino decay
can only produce sub-leading effects.

Status of the neutrino spin-flip in the magnetic field of the Sun as well
as   non-standard neutrino interactions 
is not yet clear. These mechanisms can be  
leading or produce  sub-leading effects, if {\it e.g.},  
LMA MSW turns out to be the solution.
Also  a possibility of the hybrid solutions
is not excluded when two (or more) different effects
give comparable contributions to the conversion:
for instance,  MSW conversion and the spin-flip \cite{minnun,valen-sub}.

\subsection{Solar neutrinos and  $U_{e3}$}

Let us consider the neutrino mass spectrum  
which explains the solar and  atmospheric neutrino data. 
The mass eigenstates $\nu_1$ and $\nu_2$ are split by
the solar $\Delta m^2_{12}$, whereas 
the third mass eigenstate, $\nu_3$, is separated by  mass 
difference  determined by  the atmospheric $\Delta m^2_{13}$. 
Matter effect influences  mixing (flavor content)  
of  the third mass eigenstate very weakly. 
The effect of  third neutrino is reduced then to averaged vacuum
oscillations, and  the  probability equals 
\be
P_{ee} = \cos^4 \theta_{13} P_{ee}^{(2)} +  \sin^4 \theta_{13}. 
\label{surv3}
\ee
Here $\sin \theta_{13} \equiv U_{e3}$,   
$P_{ee}^{(2)}$ is the two neutrino oscillation probability  characterized
by  $\tan^2 \theta_{12}$, $\Delta m_{12}^2$ and the effective
matter potential reduced by factor $\cos^2 \theta_{13}$ 
(see {\it e.g.} \cite{three1,Fogli3nu} for related studies).

\null
\vskip 0.7cm  
\begin{figure}[htb]
\centering\leavevmode
\epsfxsize=.7\hsize
\includegraphics[width=0.52\textwidth]{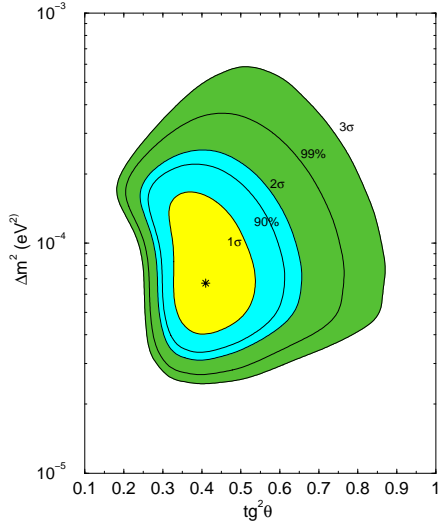}
\vskip -1.5cm
\caption{
Global LMA solution for 
$\sin^2\theta_{13}=0.04$.  
The best fit point is marked by a star.  
}
\label{globallmaue3} 
\end{figure}
Mainly, the effect of $\theta_{13}$  consists of  overall
suppression of the survival probability. The  suppression  
factor can be as small as 0.90 - 0.92. 
Results of the global analysis in the three neutrino context are shown 
in fig.~\ref{globallmaue3}.
We use the three neutrino survival probability (\ref{surv3}) for fixed
value $\sin^2 \theta_{13} = 0.04$ (near the upper bound from the 
CHOOZ experiment \cite{CHOOZ}), so that number of degrees of freedom
is the same as in the previous analysis.
In the best fit point:
\be
\Delta m^2_{12} = 6.7 \cdot 10^{-5} {\rm eV}^2 , ~~  
\tan^2\theta  = 0.41, 
\label{global3nu}
\ee
and $f_B = 1.09$.  
The best fit value of $\Delta m^2_{12}$ is
slightly higher than that  in the two neutrino case.  
The changes are rather small, however, as a tendency, 
with increase of $\theta_{13}$ the fit becomes worse 
in comparison with $2\nu-$ case: 
for $\sin^2 \theta_{13} = 0.04$ we get  $\Delta \chi^2 = 1.0$. 
The detailed study of the conversion in $3\nu-$ context is given 
in~\cite{Fogli3nu}. 

For LOW solution increase of $\theta_{13}$ leads to improvement of the
fit, so that this solution appears (for $\sin^2\theta_{13}=0.4$) at
3$\sigma$ level with respect to best fit point (\ref{global3nu}).

\subsection{Sterile neutrinos}

\begin{figure}[ht]
\centering\leavevmode
\epsfxsize=.8\hsize
\epsfbox{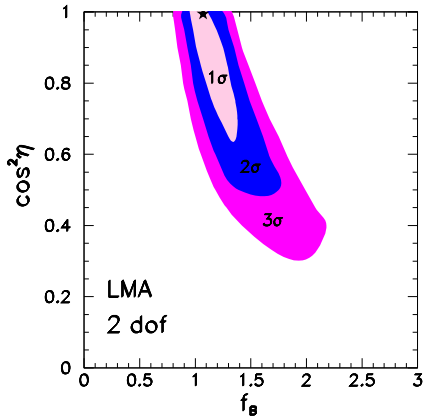}
\vskip -1cm
\caption{Bounds on sterile neutrino admixture and the original 
boron neutrino flux, (from \cite{sterile}). 
}
\label{ster}
\end{figure}

Oscillations to pure sterile state: $\nu_e  \rightarrow \nu_s$ are
excluded at $5\sigma$ level \cite{sno-nc}. So, the  sterile
neutrinos, if exist,  may produce  a sub-leading effect, 
or be a part of ``hybrid" solution.  
The analysis  has been performed in
the context
of  single $\Delta m^2$ conversion  $\nu_e
\rightarrow \nu_x$,
where
\be      
\nu_x = \cos \eta~ \nu_a + \sin \eta ~\nu_s. 
\ee
In this case $\nu_e$ undergo  transitions to $\nu_e,
\nu_{\mu,\tau}, \nu_s$
inside the Sun with the following probabilities:
\be
\nu_e \rightarrow \left\{
\begin{array}{ll}
\nu_e, & P_{ee}\\
\nu_{a}, & (1 - P_{ee}) \cos^2\eta \\
\nu_{s}, & (1 - P_{ee}) \sin^2\eta \\
\end{array}.
\right.
\label{components}
\ee
The matter potential is modified:
$V = \sqrt{2} G_F(n_e - 0.5 \sin^2 \eta n_n)$, where
$n_e$ and $n_n$ are the densities of the electrons and neutrons 
correspondingly.
According to (\ref{components})
fluxes of neutrinos detected by the charged current (CC), 
neutral current 
reaction (NC) and the  neutrino-electron scattering (ES)  equal
respectively:
\be
\begin{array}{l}
\Phi_{CC} = f_B P_{ee}\\
\Phi_{NC} = f_B [1 -  (1 - P_{ee}) \sin^2\eta] \\
\Phi_{ES} = f_B [P_{ee} - r (1 - P_{ee}) \cos^2\eta] \\
\end{array}.
\label{fluxes}
\ee
The fluxes (\ref{fluxes}) depend on two combinations
of three parameters: $f_B P_{ee}$ and $f_B (1 - P_{ee}) \cos^2\eta$.
So, degeneracy of parameters takes place and increase of  
$\eta$ can be compensated by
changes  of $f_B$ and $P_{ee}$ \cite{barger-a}. The degeneracy is
broken by  energy dependence of the probability and by
the Earth regeneration effect. Since both effects 
are small (according to the data), only weak bound 
on admixture of sterile neutrino follows from the present data     
(see fig.~\ref{ster} from 
\cite{sterile}). In the analysis \cite{sterile} for each pair of values of
($\cos^2\eta$, $f_B$),
$\chi^2$ is minimized with respect to $\Delta m^2, \tan^2 \theta$, thus
function $\chi^2 (\cos^2\eta, f_B)$ has been found which was used
to construct the confidence level contours in fig.~\ref{ster}.
The absolute minimum of $\chi^2$ corresponds to no sterile neutrino:
$\sin^2\eta = 0, ~~~ f_B = 1.07$, 
and the upper  bound on  $\eta$ is rather weak
\be
\sin^2\eta < 0.35~~ (0.70), ~~~~ 1\sigma~~ (3\sigma).
\ee
Stronger bound can be obtained from  combined analysis of the
solar neutrino data and the KamLAND results.
\subsection{Spin-flavor precession}

Resonance spin-flavor precession \cite{LMA} opens unique
possibility to reconcile strong suppression of solar
neutrino flux at intermediate energies and
absence of  distortion at high energies.
For appropriate choice of the magnetic field distribution   
the mechanism allows to reproduce the profile of fig.~\ref{fi1}. 
Analysis of the data gives 
values of $\Delta m^2$ and  magnetic
field in the peak~\cite{gago-ap,gago,mmczrad,valen-sf} :
\be
\begin{array}{l}
\Delta m^2 = (0.8 - 1.5) \cdot 10^{-8} {\rm eV}^2,\\ 
B = 80 - 100 {\rm kG}
\end{array}
\label{rsfp}
\ee 
for the neutrino magnetic moment
$\mu_{\nu} =10^{-11} \mu_B$. 
Solutions with larger values of $B$ also exist.

Non-resonance spin-flavor precession leads to  profile similar to
that of  LMA MSW~\cite{miranda}.
The required values of parameters are as in ~(\ref{rsfp}).

No oscillation effect is expected in the KamLAND experiment if
one of the spin-flip solutions is true. Furthermore,  
solutions based on the neutrino spin-flip do not lead to the Earth
regeneration effects, in particular,  to the Day-Night asymmetry. 

On the other hand, long period  variations of signals
are generic consequences of the neutrino spin-flip 
in the convective zone of the Sun. 
Indeed, the surface magnetic field shows 11 years variations. 
There is no model which has constant large scale  field in the 
convective zone and variations of the surface field \cite{centf}. 
Due to  convective mixing large scale field in the convective zone should
be seen at the surface \cite{centf}.

Another variability is expected due to existence of the
equatorial gap in the toroidal magnetic field and inclination 
of the Earth  orbit with respect to solar equatorial plane. 
This leads to the semiannual variations of the electron 
neutrino flux~\cite{old}.   

Neither 11 years nor semiannual variations of signals  
have been observed. 
That was the motivation to revisit a possibility of the neutrino spin-flip 
in the radiative zone \cite{centf,mmczrad}. It was assumed  that strong  
relic field  exists in the radiative zone. The field is frozen  
and therefore constant. The field has a toroidal configuration 
with  maximal strength $B = (0.4 - 0.6)$ MG  at $R \sim 0.15 R_{\odot}$. 
The survival probability is given by 
\be
P_{ee} = 0.5 [1 - (1 - 2P_c)\cos 2\theta_0], 
\ee
where $P_c$ is the level crossing probability and $\theta_0$ is the 
mixing angle in the production point. 
To describe  experimental data one needs to have $P_{ee} \sim 0.3$, 
that is, the flip should be non-adiabatic. Notice that in the 
non-adiabatic case  it is rather non-trivial 
to have weak energy distortion of the spectrum. 

The profile of the effect is similar to the profile for LMA, 
and the data are well described for 
$\Delta m^2 = (1.5 - 5.0) 10^{-6}$ eV$^2$ \cite{centf}. 

\subsection{Non-standard neutrino interactions} 

Both mixing  and level splitting required for 
neutrino conversion can be  induced by the neutrino interaction with
matter \cite{fcnc}.  
The corresponding terms in the effective Hamiltonian
(mixing, $\bar{H}$, level splitting, $H$) can be parameterized in the
following way: 
\be
\begin{array}{l}
H = \sqrt{2} G_F [n_e(r) - \epsilon' n_f(r)]\\
\bar{H} = \sqrt{2} G_F [\epsilon n_f(r)], 
\end{array}
\ee
where $f = u, d$ (-quarks). There is no dependence of the probability 
on neutrino energy: $P_{ee} = const$. The dependence of the effect on 
energy  appears due to  difference of the production regions for 
$pp-$, $Be-$ and $B-$neutrinos. The average densities in the production
region  satisfy inequalities 
$n(pp) < n(Be) < n(B)$. The larger the density the stronger
conversion. In such a way the LMA type profile of the effect is well
reproduced. Global fit of the data leads to the  best  
values of parameters~\cite{gago}: 
\be
\epsilon,~~ \epsilon' = 
\left\{
\begin{array}{lll}
3 \cdot 10^{-3},  & 0.60, & f = u \\
1.5 \cdot 10^{-3}, &  0.43, & f = d
\end{array}
\right. . 
\ee
The problem here  is that very large value of the diagonal 
coupling $\epsilon'$ is required which is difficult to reconcile with 
the experimental bounds  in the context of known
models for $\epsilon,  \epsilon'$.  
The solution predicts earth regeneration effect, however no oscillation 
signal in KamLAND is expected. 
 
\section{CONCLUSION}

LMA MSW  with parameters $\Delta m^2
\sim (5 - 7) \cdot 10^{-5}$ eV$^2$ and $\tan^2 \theta = 0.35 - 0.45$
is rather  plausible solution.  
It fits well experimental data and  our new theoretical  
prejudices. In this sense, the present status of LMA is 
similar to the status of  SMA  5 years ago... 
So, some surprises are not excluded. 

What could be alternative? 
The  QVO-VO and  LOW solutions are  
accepted at about $3\sigma$ level. 
SMA  is excluded at more than  $5\sigma$ level. 
Solutions based on neutrino spin-flip in the solar magnetic 
field or non-standard neutrino interactions (flavor changing and flavor 
diagonal) are rather appealing from the point of view 
of the data fit, but they have their own disadvantages. 
Hopefully, the situation will be clarified by  KamLAND, further operation
of SNO and, later,  BOREXINO~\cite{bor}. 

In the context of LMA MSW, (as well as LOW) 
recent  SNO results  lead to important bounds on neutrino parameters: 
Now we have strong evidence
that ``solar'' mixing is non-maximal, and
moreover, deviation from maximal mixing is rather large:  
$\tan^2 \theta < 0.8$ ($3\sigma$). 
Possible effect of $U_{e3}$ is  small being disfavored by 
data.  The admixture of sterile component is further restricted:
$\sin^2 \eta < 0.35 - 0.50$. 

The LMA MSW solution will be tested soon  by the KamLAND experiment:
in the best fit point one expects the suppression factor for 
integral  signal $\sim 0.6 - 0.7$  (0.3 - 0.75 at $3\sigma$) and the spectrum
distortion with substantial suppression in the high energy part. 

Soon,  solar neutrino studies can be transformed  to 
new stage when emphasis will be on 
determination  of the original neutrino fluxes and 
searches for the sub-dominant  
effects produced by  $U_{e3}$, admixture of sterile neutrino both in one and 
more than one $\Delta m^2$ context, magnetic moment of neutrino,
non-standard interactions, etc.. 




\begin{thebibliography}{99}


\bibitem{sno} Q. R. Ahmad {\it et al.}, SNO collaboration,
Phys. Rev. Lett. 87:071301, (2001).
 
\bibitem{sno-nc} Q. R. Ahmad {\it et al.}, SNO collaboration,
nucl-ex/0204008.

\bibitem{sno-dn} Q. R. Ahmad {\it et al.}, SNO collaboration,
nucl-ex/0204009.
 
\bibitem{sno-how} ``How to use the SNO Solar Neutrino Spectral Data'', at 
http://www.sno.phy.queensu.ca/.

\bibitem{Cl} B. T. Cleveland  {\it et al.,} Astroph. J. {\bf 496}
(1998) 505. 


\bibitem{SK} S. Fukuda {\it et al.} (Super-Kamiokande collaboration) 
Phys. Rev. Lett. 86: 5651, 2001;  Phys. Rev. Lett. 86: 5656, 2001.   


\bibitem{sk-nu} M.B. Smy, these proceedings.    
 

\bibitem{sage} J.N. Abdurashitov et al.,  
astro-ph/0204245, V. Gavrin, these proceedings.    

\bibitem{gno} T. Kirsten, these proceedings



\bibitem{BL} V. S. Berezinsky, M. Lissia, hep-ph/0108108.

\bibitem{FLMPa} G. L. Fogli et al., hep-ph/0203138.

\bibitem{barger-a} V. Barger, D. Marfatia, K. Whisnant, B. Wood,
hep-ph/0204253. 


\bibitem{BB} J. N. Bahcall, H. Bethe,
Phys. Rev. Lett. 65:2233, 1990.

\bibitem{Ber} V. S. Berezinsky, 
Comments Nucl. Part. Phys. 21:249, 1994.  


\bibitem{Stur} P. A. Sturrok, M. A. Webber,
Astrophys. J., 565, 1366 (2002); 
P. A. Sturrok,  J. D. Scargle, Astrophys. J., 550, L101 (2001).
   
\bibitem{BKS10} J. N. Bahcall, P. I. Krastev, 
A. Yu. Smirnov,  Phys. Rev. D62:093004, 2000.  



\bibitem{bahcall-a}John N. Bahcall, M.C. Gonzalez-Garcia, Carlos
Pe\~na-Garay,  hep-ph/0204314.  

\bibitem{dproy} A. Bandyopadhyay, S. Choubey,
S. Goswami, D. P. Roy, hep-ph/0204286.  

\bibitem{strumia} P. Creminelli, G. Signorelli 
A. Strumia, hep-ph/0102234, v3 22 April 2002 (addendum 2). 

\bibitem{milano} P. Aliani, et al, hep-ph/0205053. 

\bibitem{pedro} P. de Holanda, A. Yu. Smirnov, hep-ph/0205241, v3.  

\bibitem{cattadori} A. Strumia, C. Cattadori, N. Ferrari, F. Vissani,
hep-ph/0205261.

\bibitem{bari} G.L. Fogli, E. Lisi, A. Marrone,
D. Montanino, A. Palazzo, hep-ph/0206162.  
 
\bibitem{valen-so}M. Maltoni, T. Schwetz, M. A. Tortola, J. W. F. Valle,
hep-ph/0207227.  



\bibitem{ssm} J. N. Bahcall, M.H. Pinsonneault and S. Basu,  
Astrophys. J. {\bf 555} (2001)990.  

\bibitem{hepfl} 
L. E. Marcucci {\it et al.}, Phys. Rev. {\bf C63} (2001)  015801;   
T.-S. Park, {\it et al.}, hep-ph/0107012 and references 
therein. 


\bibitem{max} M. C. Gonzalez-Garcia, C. Pe\~na-Garay, Y. Nir,
A. Yu. Smirnov,  Phys. Rev. {\bf D63} (2001) 013007. 


\bibitem{CHOOZ} CHOOZ Collaboration, M. Apollonio {\it et al.},
Phys.Lett. {\bf B420}, 397 (1998).

 
\bibitem{lma99} J. N. Bahcall, P.I. Krastev, A. Yu. Smirnov, Phys. Rev. 
{\bf D60} (1999) 093001. 

\bibitem{zenith} M. C. Gonzalez-Garcia,  C. Pe\~na-Garay,  A. Yu. Smirnov, 
Phys. Rev. {\bf D63} (2001) 113004.



\bibitem{busenitz}
P. Alivisatos et al., KamLAND, Stanford-HEP-98-03, Tohoku-RCNS-98-15. 

\bibitem{plamen}P. I. Krastev, A.Yu. Smirnov, Phys. Rev. D65:073022, 2002.  

\bibitem{minnun} H. Minakata, H. Nunokawa 
Phys. Rev. Lett. 63:121, 1989.   



\bibitem{valen-sub} W. Grimus, M. Maltoni,
T. Schwetz,  M.A. Tortola, J.W.F. Valle, hep-ph/0208132.  



\bibitem{three1}A. Bandyopadhyay, S. Choubey, S. 
Goswami, Kamales Kar, Phys. Rev. D65:073031, 2002.     

\bibitem{Fogli3nu}G. L. Fogli et al., hep-ph/0208026. 

\bibitem{sterile}J. N. Bahcall, M. C. Gonzalez-Garcia, C. Pe\~na-Garay, 
hep-ph/0204194.


\bibitem{LMA} C.S. Lim,  W. J. Marciano, Phys. Rev. D37 1369 (1988); 
E. Kh Akhmedov, Phys. Lett. B213, 64 (1988).  

\bibitem{gago-ap} E. K. Akhmedov, J. Pulido,  
Phys. Lett. B529: 193 (2002). 


\bibitem{gago} A. M. Gago, et al, Phys. Rev. D65:073012, 2002. 

\bibitem{mmczrad}B. C. Chauhan, J. Pulido
hep-ph/0206193.   

\bibitem{valen-sf} J. Barranco, O.G. Miranda, T.I. Rashba, V.B. Semikoz, 
J.W.F. Valle, hep-ph/0207326. 

\bibitem{miranda} O. G. Miranda et al., hep-ph/0108145. 


\bibitem{centf} A. Friedland, A. Gruzinov, hep-ph/0202095.  

\bibitem{old}A. I. Veselov, M.I. Vysotsky, V.P. Yurov 
Yad.Fiz. 45:1392, 1987;  
P.I. Krastev, A.Yu. Smirnov, Z. Phys. C49:675, 1991.   



\bibitem{fcnc} L. Wolfenstein, Phys. Rev. {\bf D17} 2369 (1978); 
E.  Roulet, Phys. Rev. {\bf D44} R935 (1991); 
M. M. Guzzo, A. Masiero, S. T. Petcov, Phys. Lett. B260:154 (1991).



\bibitem{bor} G. Bellini,  for BOREXINO Collaboration,  these proceedings. 

\end{thebibliography}
\end{document}